\newcommand{\bfx}{\mbox{\boldmath $x$}}
\newcommand{\bfz}{\mbox{\boldmath $0$}}

\newcommand{\bfU}{\mbox{\boldmath $\hat{U}$}}
\newcommand{\bfplus}{\mbox{\boldmath $+$}}
\newcommand{\bfminus}{\mbox{\boldmath $-$}}
\newcommand{\bfsigma}{\mbox{\boldmath $\sigma$}}

\documentclass[aps,pre,twocolumn,showpacs]{revtex4}
\usepackage{graphicx,amssymb}
\begin{document}
\title{Constrained spin dynamics description of random walks on
hierarchical scale-free networks}
\author{Jae Dong Noh}
\affiliation{Department of physics, Chungnam National University, 
Daejeon 305-764, Korea}
\author{Heiko Rieger}
\affiliation{Theoretische Physik, Universit\"at des Saarlandes, 
66041 Saarbr\"ucken, Germany}

\begin{abstract}
We study a random walk problem on the hierarchical network which is a
scale-free network grown deterministically. 
The random walk problem is mapped onto a dynamical Ising
spin chain system in one dimension with a nonlocal spin update rule, which
allows an analytic approach.
We show analytically that the characteristic relaxation 
time scale grows algebraically with the total number of nodes $N$ as 
$T \sim N^z$. From a scaling argument, we also show 
the power-law decay of the autocorrelation function $C_{\bfsigma}(t)\sim
t^{-\alpha}$, which is the probability to find the Ising spins 
in the initial state ${\bfsigma}$ after $t$ time steps, with the
state-dependent non-universal exponent $\alpha$. It turns out that the
power-law scaling behavior has its origin in an quasi-ultrametric
structure of the configuration space.
\end{abstract}
% 05.40.Fb : Random walks and Levy flights
% 89.75.Hc : Networks and genealogical trees
% 05.60.-k : Transport processes

\maketitle
\section{Introduction}\label{sec1}
Complex networks, as for instance represented by the Internet, the
social acquaintance network between individuals, biological networks
of interacting proteins, and others (see Ref.~\cite{Rev_articles} for
further examples), became recently a central research focus in
statistical physics.  In general a network consists of a set of
nodes~(sites or vertices) and a set of edges~(bonds or arcs),
connecting the nodes with one another.  A system with many interacting
degrees of freedom, e.g., computers, individuals, proteins etc., or
generally called agents, can be modeled by a network by identifying
the agents as the nodes and the interaction between them as the edges.
Real world networks neither have a regular structure (as for instance
periodic lattices or grid graphs have) nor a fully random structure
\cite{Watts98}. They rather display a broad distribution of the degree,
where the degree $K$ of a node is the number of neighbors connected to
it. Some networks, the so-called scale-free networks \cite{AHB99},
display a power-law degree distribution $P(K) \sim K^{-\gamma}$, which
is is found in various disciplines.

The heterogeneous structure of scale-free networks has a significant
influence on thermodynamic or dynamic systems embedded into them. For
instance, the nature of equilibrium~\cite{Etr} or
nonequilibrium~\cite{NonEtr} phase transitions are quite different
from those observed in corresponding systems on regular periodic
lattices. In the present work we are interested in the nature of
diffusive and relaxational dynamics performed by a random walker in
scale-free hierarchical network~\cite{Ravasz&Barabasi03}. As a very
recent application we note that in the context of peer-to-peer
computer networks random walk search strategies have been proposed
\cite{adamic,lv,sarshar}, in which a query message is forwarded to a randomly
chosen neighbor at each step until the desired object (typically a
particular data set) is found.  In view of these algorithmic
developments it appears therefore quite natural and important to study
random walks on complex networks. In addition, the random walk is a
fundamental stochastic process~\cite{Hughes95} and turns out to be a
useful tool in characterizing the structure of complex
networks~\cite{Newman,Noh&Rieger03,Zhou}.

In regular networks of periodic lattices in $D$ dimension, the random
walk motion is characterized by normal diffusion which is
characterized by a length scale that grows algebraically as $\xi
\sim t^{1/2}$ in time $t$. The exponent $1/2$ is universal, i.e.\ 
it does not depend on the microscopic details of the lattice --- the
only condition being that only nearest neighbor jumps on a regular
$D$-dimensional lattice are allowed. The autocorrelation function
$C(t)$ or the return probability to the initial node in $t$ time
steps decays algebraically as $C(t) \sim t^{-D/2}$. On random
networks, on the other hand, the autocorrelation function shows a
stretched-exponential decay as $C(t) \sim e^{-a t^\beta}$ with $\beta
= 1/3$~\cite{Bray&Rodgers88}. 

Random walks were also studied in the small-world network of Watts
and Strogatz~\cite{Watts98}, which interpolates between regular networks
and random networks by stochastically changing connections between nodes 
with a particular rewiring probability $p_W$.
In essence a small-world network is obtained from a regular
network with edges of fraction $p_W$ being replaced by shortcuts
connecting pairs of nodes selected randomly. For nonzero $p_W$, an
interesting crossover behavior is observed~\cite{Pandit,AlmaasKS03}: A
random walk obeys the scaling law for regular networks for short
times $t\ll \tau$, and then that for the random networks for large
times $t\gg \tau$.  The crossover time scale $\tau$ is determined by
the time interval at which a random walker hits shortcuts. Since the
mean distance between shortcuts is $\xi \sim p_{W}^{-1}$, the
crossover time scales as $\tau
\sim \xi^2 \sim p_W^{-2}$. For $t\gg \tau$, it is numerically found that
the autocorrelation function also shows a stretched-exponential decay 
as $C(t) \sim e^{-a t^{\beta}}$ with $\beta\simeq
1/3$~\cite{jespersen,lahtinen}.

There has been a growing interest recently in the study of random
walks on scale-free networks~\cite{Tadic1,Tadic2}.  In this paper, we
focus on random walks on a hierarchical network, which is a model for
a scale-free network with a modular
structure~\cite{Ravasz&Barabasi03}.  Unlike most scale-free network
models it is a deterministic network as those of Jung {\em et
al.}~\cite{JungKK02} and Dorogovtsev {\em et
al.}~\cite{DorogovtsevGM02}. Due to its deterministic nature a number
of characteristic structural features are known exactly~\cite{Noh03}.
As we will see in the following, we can study various properties of
the random walk analytically. The analytic results will shed light on
the stochastic processes in general scale-free networks.

The paper is organized as follows: In Sec.~\ref{sec2}, the
hierarchical network model and the random walk is introduced. Our
results for the scaling laws for the relaxation time and the
autocorrelation functions are presented in Sec.~\ref{sec3}. These
results are derived with the help of an exact mapping of the random walk
problem to a constrained dynamics of an Ising spin chain, the details of which
mapping are described in Sec.~\ref{sec4}. We also find that a random
walk on a hierarchical network is similar to the diffusion in
ultrametric space, which is elaborated in Sec.~\ref{sec5}. Finally we
summarize our work in Sec.~\ref{sec6}.

\section{Model}\label{sec2}
Some biological networks which are scale-free exhibit a modular
structure, which is not incorporated into most scale-free network
models. The hierarchical network has been proposed as a model for the
scale-free networks with the modular
structure~\cite{Ravasz&Barabasi03}. It is constructed iteratively
starting from a seed~(first generation) ${\cal G}_1$ consisting of a
{\em hub} and $(M-1)$ {\em peripheral nodes}. They are fully connected
with each other. It is useful to represent the hub and the peripheral
nodes with the coordinates $(0)$ and $(y)$, where $y$ is an integer 
$1\leq y<M$~\cite{Noh03}. Nodes in ${\cal G}_G$, the network of the
$G$th generation, are identified via coordinates that are $G$-tuples 
of integers $(\bfx)=(x_G\ldots x_1)$.

\begin{figure}
\includegraphics*[width=\columnwidth]{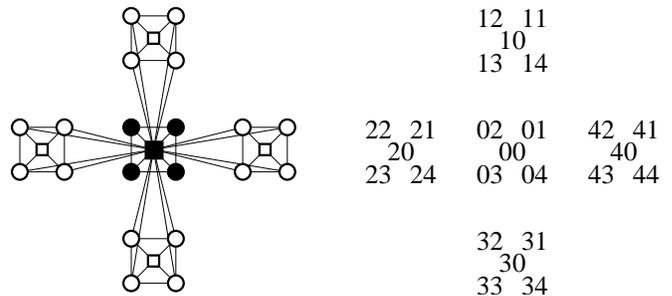}
\caption{The configuration and the coordinate representation
of ${\cal G}_2$ with $M=5$. The hub is represented with a  
filled square, and the peripheral nodes are with empty circles.}\label{fig1}
\end{figure}

From a given graph ${\cal G}_g$, the next generation network ${\cal
G}_{g+1}$ is constructed by adding $(M-1)$ copies of ${\cal G}_{g}$ with
their peripheral nodes connected to the hub of the original ${\cal
G}_{g}$. The original hub and the peripheral nodes in the copies
become the hub and peripheral nodes of ${\cal G}_{g+1}$, respectively.
Then, each node whose coordinate was $(\bfx)$ is assigned to $(0\bfx)$
if it belongs to the original ${\cal G}_{g}$ or to $(y\bfx)$ with
$1\leq y<M$ if it belongs to the $y$th copy of ${\cal G}_{g}$.  So the
hub has $x_n=0$ for all $n$ and a peripheral node has $x_n\neq 0$
for all $n$. There are $M^G$ nodes in ${\cal G}_G$, $(M-1)^G$ of which
are peripheral nodes.  Figure~\ref{fig1} shows the configuration and
the coordinate representation of ${\cal G}_2$ with $M=5$.  The
iteration can be repeated indefinitely and the emerging network is
scale-free for $M\geq 3$ with the degree distribution exponent $\gamma
= 1 +
\ln M / \ln (M-1)$~\cite{Noh03}. 

The node connectivity is represented by the adjacency matrix $A_{ji}$;
$A_{ji}=1$ if a node $i$ is connected to $j$ or 0 otherwise. The network is
undirected, hence $A_{ij} = A_{ji}$ and the connectivity is
easily described in terms of the coordinates~\cite{Noh03}.
Hereafter, we will use $x$ for a dummy index from $0$ to $M-1$, while $y$
from $1$ to $M-1$, and we denote the $m$-tuple of $0$ as $\bfz_m$. 

The network growth rule implies {\bf (a)} the existence of connections
of $m$-th generation hub to all $m$-th generation peripheral nodes,
more precisely in coordinate language: nodes $(\bfx)$ with $x_i=0$ for
$i=1,\ldots,m$ and $x_{m+1}=y_{m+1}\neq 0$ are connected to the
following nodes:
\begin{equation}\label{z2nz}
(\cdots y_{m+1}\bfz_m) \leftrightarrow 
(\cdots y_{m+1}\bfz_{m-n} y_n \cdots y_1)
\end{equation}
with $1\leq n \leq m$. And it implies {\bf (b)} the existence of
connections between peripheral nodes and lower level hubs plus
connections to other peripheral nodes within the same elementary unit;
in coordinate language: a node $(\bfx)$ with $x_i=y_i\neq 0$ for
$i=1,\ldots,m$ and $x_{m+1}=0$ is connected to the following nodes:
\begin{equation}\label{nz2z}
(\cdots 0y_m\cdots y_1) \leftrightarrow  \left\{
\begin{array}{l}
(\cdots 0y_m\cdots y_2 y'_1) \\[3mm]
(\cdots 0y_m \cdots y_{n+1} \bfz_n)
\end{array} \right.
\end{equation}
with $y'_1\neq y_1$ and $1\leq n\leq m$.

We study a discrete time random walk on the network.  This stochastic
process is defined by the following rules: The walker at node $i$ and
time $t$ selects one of the neighbors of $i$ to which $i$ is connected
and jumps to this neighbor at time $t+1$.  Thus the transition
probability for a jump from a node $i$ to a node $j$ is given by
$\omega_{ji} = A_{ji}/K_i$, where $A_{ji}$ is the adjacency matrix and
$K_i = \sum_j A_{ji}$ is the degree of the node $i$.

This stochastic process in discrete time is described by a master
equation for the time evolution of $P_i(t)$, the probability finding
the walker at node $i$ and time $t$. The master equation reads
$P_i(t+1) = \sum_j \omega_{ij}P_j(t)$.  Equivalently, defining the
state vector $|P(t)\rangle \equiv \sum_i P_i(t) |i\rangle$ with
$|i\rangle$ being the state in which the walker is at node $i$, one
can rewrite the master equation as $|P(t+1)\rangle = \bfU
|P(t)\rangle$, where $\bfU$ is the transition operator whose elements
are $(\bfU)_{ji} = \omega_{ji}$.

In the infinite time limit $t\to\infty$ the probability distribution
converges to the stationary state distribution $P^\infty_i$, which is
given by $P_i^\infty = K_i / {\cal N}$ with ${\cal N} \equiv \sum_i
K_i$ for the random walk on arbitrary undirected
network~\cite{Noh&Rieger03}.  In the hierarchical network the degree
of all nodes are known exactly~\cite{Noh03}.  For instance, the hub
has the largest degree
\begin{equation}\label{K_h}
K_{h} = (M-1)(M-2)^{-1} ((M-1)^G-1)\sim (M-1)^G \ ,
\end{equation} 
and the peripheral node has the degree 
\begin{equation}\label{K_p}
K_p =(M-2+G) \ .
\end{equation}
The sum of all degrees is given by
\begin{equation}\label{calN}
{\cal N} = (3M-2)(M-1)M^{G-1} - 2 (M-1)^{G+1} \sim M^G \ .
\end{equation}

A quantity of particular interest is the scaling law for the
relaxation time $T$, which is the characteristic time scale for the
approach of the probability distribution $P_i(t)$ to the stationary
state distribution $P_i^\infty$. Also of interest is the nature of the
relaxation dynamics, for which we consider the decay of the
autocorrelation function
\begin{equation}\label{C_def}
C_{S}(t) = \langle S | {\bfU}^t | S \rangle \ ,
\end{equation}
which is the overlap between a state $|S\rangle$ with itself after $t$
time steps. When $|S\rangle = |i\rangle$, it reduces to the returning
probability of the random walker to the origin~(starting node) $i$
after $t$ time steps. In the limit $t\to\infty$ the autocorrelation
function converges to a value determined by the stationary state
distribution $P^\infty$. The scaling behavior of $C_{S}(t)$ for 
$t\ll T$ will be studied for various states $|S\rangle$.

\section{Results}\label{sec3}
In this section we present our the main results.  They are derived
using the exact mapping of our random walk process onto a constrained 
dynamics of an Ising spin chain.  Details of the mapping and the
derivations of the formulas deduced from it and used in the present
section are delegated to the next section.

\subsection{Relaxation time}
Consider the motion of the random walker located initially on a
particular node, say $(030201)$. The memory of the initial
position will be lost when all components $x_i$'s are flipped at least
once, which defines the relaxation time scale $T$. The node
connectivity summarized in Eqs.~(\ref{z2nz}) and (\ref{nz2z}) tells us
that $x_i$ may flip only when all $x_j$'s with $j<i$ are equal to
$0$~(if $x_i=0$) or all are not equal to $0$~(if $x_i\neq 0$). Hence,
the random walker should follow the path $(0302 01) \rightarrow
(0302 00) \rightarrow (0302 y_2 y_1) \rightarrow (0300 00) \rightarrow
(03 y_4 y_3 y_2'y_1') \rightarrow (0000 00) \rightarrow (y_6 y_5 y_4'
y_3' y_2'' y_1'')$ to loose the memory of its initial state.

Each process requires a simultaneous flip of $\xi$ components from
zero to nonzero values or vice versa, which may occur after many
trials.  For instance, the random walker at $(0302y_2y_1)$ may hop to
$(0302y_2 0)$ or $(030200)$ instead of to $(030000)$.  When it jumps
to a wrong node, say $(030200)$, first it should hop to a node
$(0302y_2'y_1')$, and then try another hopping toward the
destination. In this respect the dynamics we are considering is of a
hierarchical nature. Utilizing this observation we will show in
the next section that the associated time scale $\tau_\xi$ for the
process increases exponentially as $\tau_\xi \sim (M/(M-1))^\xi$. We
define $\kappa \equiv M/(M-1)$ for further use.

Therefore, the relaxation time $T$, which is given by 
$T \sim \sum_{\xi}^G \tau_\xi$, scales {\em exponentially} 
with $G$ as 
\begin{equation}\label{T_G}
T \sim \kappa^G \ .
\end{equation}
Since $N = M^G$, the relaxation time scales {\em algebraically} with $N$ as
\begin{equation}\label{T_N}
T \sim N^z \ , 
\end{equation}
with the dynamic exponent
\begin{equation}\label{exp_dynamic}
z = \ln \kappa  / \ln M \ .
\end{equation}

\subsection{Autocorrelation function}
To be specific we consider the autocorrelation functions for the
following states:
\begin{enumerate}
\item[(i)] $H$ is the state corresponding to the hub
\begin{equation}\label{Hstate}
|H \rangle = | \bfz_G \rangle \ .
\end{equation}
\item[(ii)] $P$ is the state corresponding to the peripheral nodes
\begin{equation}\label{Pstate}
|P\rangle = \frac{1}{(M-1)^G} \sum_{y_1\ldots y_G} | y_G \ldots
y_1\rangle \ .
\end{equation}
\item[(iii)] $A1$ and $A2$ are the states 
\begin{eqnarray}
|A1\rangle &=& \frac{1}{(M-1)^{(G/2)}} \sum_{y_2, y_4, \ldots} | \ldots
y_40y_30\rangle \ , \label{A1state} \\ 
|A2\rangle &=& \frac{1}{(M-1)^{(G/2)}} \sum_{y_1, y_3, \ldots} | \ldots
0y_30y_1\rangle \label{A2state}
\end{eqnarray}
with zero and nonzero components alternating.
\end{enumerate}

The stationary state probability distribution is determined by the degree
distribution. Since the degree of all nodes is known, it is easy
to show that $P_H^\infty \sim \kappa^{-G}$, $P_P^\infty \sim G \kappa^{-G}$,
and $P_{A1}^\infty = P_{A2}^\infty \sim (M/\sqrt{M-1})^{-G}$ in the large 
$G$ limit. 

The exponential decrease of the stationary state probability and
the exponential increase of the relaxation time suggests a power-law
decay of the autocorrelation function in time.
Indeed, we find that the autocorrelation functions decay algebraically
for $t\ll T$ as
\begin{equation}\label{C_H}
C_H(t) \sim \frac{1}{G^2}\ t^{-\alpha_H} 
\end{equation}
\begin{equation}\label{C_P}
C_P(t) \sim \frac{1}{G}\ t^{-\alpha_P} 
\end{equation}
\begin{equation}\label{C_A}
C_{A1}(t) \simeq C_{A2}(t) \sim t^{-\alpha_A}
\end{equation}
where $\alpha_H = \alpha_P = 1$ and $\alpha_A = \ln(M/\sqrt{M-1}) /
\ln\kappa > 1 $ with $\kappa = M / (M-1)$.
Quite remarkably, the decay exponent depends on the state --- a
manifestation of the fact that the network under consideration is not
homogeneous. In addition to the power-law dependency in $t$, the
functions $C_P(t)$ and $C_H(t)$ also decay as $1/G$ and $1/G^2$,
respectively, i.e.\ algebraically with the number of generations in
the network.  So, the power-law decay in time is observed only in
finite systems for the states $H$ and $P$, since in the limit
$G\to\infty$ the functions $C_H$ and $C_P$ vanish.

\section{Ising spin chain}\label{sec4}
In this section, we explain the exact mapping of the random walk problem 
onto the constrained dynamics of an Ising spin chain.

\subsection{Mapping}
Using the coordinate representation of the nodes, one may map the
state $|i\rangle$ with a random walker at a node $i=(\bfx)$ in ${\cal
G}_G$ to a spin configuration $|\bfx\rangle=|x_G,\ldots,x_1\rangle$
of an $M$-state Potts spin chain of length $G$, where
$x_n\in\{0,\ldots,M-1\}$ denotes the state of the spin at site $n$
($=1,\ldots,G$) in the chain.  A jump of the walker corresponds to a
transition between spin configuration. In this way the connection rules
define the time evolution of the spins.

In the context of spin dynamics, it is useful to define a {\em zero
domain}~(ZD) and a {\em nonzero domain}~(NZD); the ZD is a domain of
spins that are all in the zero-state, i.e.\
$x_i=x_{i+1}=\ldots=x_{i+l}=0$; and the NZD is one in which all spins
are in a non-zero state. In particular, a domain including the spin
$x_1$ will be called a {\em boundary domain}.  The node connectivity
imposes the constraint that spins outside the boundary domain cannot flip
in a given spin configuration.  So it suffices to consider the
transition of spins in the boundary domain. Equation~(\ref{z2nz})
implies that spins in a boundary ZD evolve in one 
time step according to $\bfU |
\bfz_m\rangle = \sum_{n=1}^m (\sum_{y_1,\cdots,y_n} |
\bfz_{m-n} y_n \cdots y_1\rangle )/\Omega$ with $\Omega =
\sum_{n=1}^m(M-1)^n$. On the other hand, Eq.~(\ref{nz2z}) implies that spins
in a boundary NZD evolve as 
$\bfU | y_m \cdots y_1 \rangle = 
(\sum_{y_1' \neq y_1} | y_m \cdots y_2 y_1'\rangle+ 
\sum_{n=1}^m | y_m \cdots y_{n+1} \bfz_n\rangle)/\Omega'$ with 
$\Omega'=(M-2+m)$.
Note that the boundary domain size decreases in most cases. 
It increases only when
all spins in the boundary domain flip.

The operator $\bfU$ is symmetric under any permutation
$y_n \rightarrow y_n'$ among nonzero spin states. Taking advantage of the
symmetry, we restrict ourselves to the subspace 
which is invariant under all such permutations. The subspace is spanned 
by the states
$|\bfsigma\rangle = |\sigma_G \cdots \sigma_1 \rangle =
|\sigma_G\rangle\otimes \cdots \otimes |\sigma_1\rangle$,
where $\sigma_n = \pm$ and 
\begin{eqnarray}
|+\rangle &\equiv& \frac{1}{M-1} \sum_{y=1}^{M-1} | y\rangle  \\
|-\rangle &\equiv& |0\rangle 
\end{eqnarray} 
For example, in ${\cal G}_2$ with $M=5$ as shown in Fig.~\ref{fig1},
$|--\rangle$ corresponds to the state with the walker at the hub, and
the states $|+-\rangle$, $|-+\rangle$, and $|++\rangle$ correspond to
the states in which the walker can be found with equal probability on
nodes $\Box$'s, $\bullet$'s, and $\circ$'s, respectively.

One may regard the two-state variable $\sigma$ as the Ising spin. 
Then the random walk 
problem in the subspace reduces to an {\bf Ising spin chain} 
with a particular constrained dynamics. 
In fact, the state defined in the previous section are equal 
to the ferromagnetically and antiferromagnetically ordered states:
\begin{eqnarray}
|H\rangle &=& |--\cdots--\rangle \label{HstateI}\\
|P\rangle &=& |++\cdots++\rangle \label{PstateI}\\
|A1\rangle &=& |\cdots+-+-\rangle \label{A1stateI}\\
|A2\rangle &=& |\cdots-+-+\rangle  \label{A2stateI}\  .
\end{eqnarray}

The Ising spins evolve as follows:
As in the Potts spin dynamics, only spins in the boundary domain may 
flip. A boundary domain $|\mbox{\boldmath $\pm$}_m\rangle$ of up/down spins 
of size $m$ evolves in one time step according to
\begin{eqnarray}
\bfU | \bfplus_m \rangle &=& \sum_{n=0}^m p_{m,n} | \bfplus_{m-n}
\bfminus_{n}\rangle \label{pmn}\\
\bfU | \bfminus_m \rangle &=& \sum_{n=1}^m q_{m,n}| \bfminus_{m-n} \bfplus_n
\rangle \ , \label{qmn}
\end{eqnarray}
where $p_{m,n} =( (M-2)\delta_{n,0} + (1-\delta_{n,0}))/(M-2+m)$ and
$q_{m,n} = (M-1)^n / \sum_{k=1}^{m}(M-1)^k$. Each spin state has a different
multiplicity factor, so the transition probabilities $p_{m,n}$ and $q_{m,n}$
are not uniform in $n$.
Note that the spin up-down symmetry is broken for $M\geq 3$.
It is restored for $M=2$, in which case the corresponding network 
is not scale-free.

\begin{figure}
\includegraphics*[width=\columnwidth]{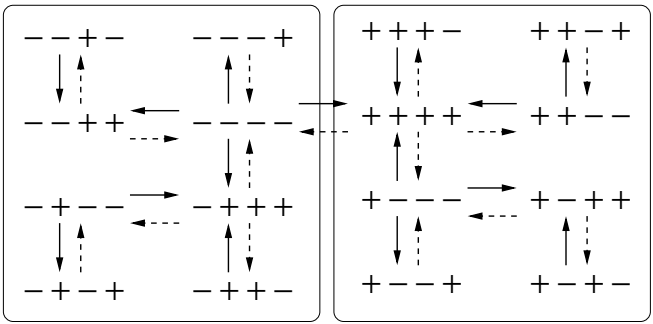}
\caption{The configuration space of the Ising spin chain of length $G=4$. 
Solid~(Dashed) lines with the arrow represent the 
transition with the probability $q_{m,n}$~($p_{m,n}$), where $m$ is the
boundary domain size of a source state and $n$ is the number of flipped
spins.  Self-loops from states with a $+$ boundary domain to themselves 
with weights $p_{m,0}$ are omitted. The parts inside the boxes with
$\sigma_4$ being ignored are equivalent to the configuration space of ${\cal
G}_3$; the configuration space has a hierarchical structure.}
\label{fig2}
\end{figure}
In Fig.~\ref{fig2}, we illustrate a diagram of the configuration space 
of the spin chain of length $G=4$ displaying the spin configurations
and possible transitions between them. Qualitative features of the Ising spin
dynamics are easily read off from the diagram:
(i) The configuration space has a tree structure, if one ignores
self-loops from the states with a $+$ boundary domain to
themselves~($p_{m,0}\neq 0$).
(ii) The configuration space has a hierarchical structure, that is, 
the configuration space of ${\cal G}_G$ contains those of ${\cal G}_{G'}$ 
with $G'<G$ as parts~(see Figure.~\ref{fig2}). 

\subsection{Boundary domain growth}\label{subsec:BDG}
The condition that only spins in a boundary domain may flip imposes 
severe constraint on the spin relaxation dynamics. 
In a given Ising spin configuration, $\sigma_2$ may flip after $\sigma_1$ 
aligns parallel to it, $\sigma_3$ may flip after $\sigma_1$ and $\sigma_2$
align parallel to it, and then, in general, $\sigma_m$ may flip after all
spins $\sigma_{n}$ with $n<m$ align parallel to it. 
In other words, the boundary domain size grows up to $m$ in order to
flip $\sigma_m$. The boundary domain growth
is the essential mechanism in the spin relaxation dynamics.
In the language of a domain wall, a domain wall at site $n+1/2$, 
i.e., $\sigma_{n+1}\neq \sigma_{n}$, plays a role of a dynamic barrier 
since it prevents spins $\sigma_{m}$ with $m>n$ from flipping.

Consider a spin configuration with a boundary domain of size $m$. 
The size of the boundary domain increases only when all $m$ 
spins inside the domain flip simultaneously. 
When $n<m$ spins flip, the boundary domain size reduces 
to $n$. Then the spin system should grow the boundary domain size up 
to $m$ to return to the initial state and try another flip to increase
the boundary domain size. It shows that the boundary growth process has
a hierarchical nature, which is inherited from to the hierarchical structure
of the configuration space.

We investigate the characteristic time scale associated with the boundary
domain growth process.  Due to the hierarchical nature of the dynamics, 
we find that the time scale satisfies a recursion relation.
To be more specific, we consider the mean first passage time~(MFPT)
$T^{+-}_m$~($T^{-+}_m$), which it takes to flip all spins in 
the boundary domain of $m$ up~(down) spins simultaneously 
for the first time. Note that such time scales do not depend on spins 
outside the boundary domain, so they do not depend on the total chain 
length $G$.

\begin{figure}
\includegraphics[width=.7\columnwidth]{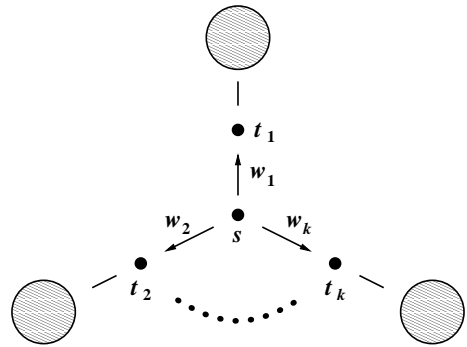}
\caption{Tree-like network. The shaded areas need not have a tree structure
as long as there is no overlap between different parts.}\label{fig3}
\end{figure}
Before proceeding, we derive a useful formula for the MFPT in a 
{\em tree-like} structure. Consider a node~(or state) 
$s$ which is connected to $k$ nodes $t_i$ with $i=1,\cdots,k$. 
The transition probability from $s$ to $t_i$ is given by $\omega_i$, and to
itself by $\omega_0$~(see Fig.~\ref{fig3}).
By the tree-like structure, we mean that $t_i$ can be reached from 
$t_j$ only through $s$ for all pairs of $i$ and $j$, no matter how many
loops there are in the shaded areas. 
Then, $T_i$, the MFPT from $s$ to $t_i$, is given by
$$
T_i = \omega_i + \sum_{j\neq i} ( 2 + T'_j ) \omega_j \omega_i \\
   + \sum_{j,j'\neq i} ( 3 + T'_j + T'_{j'}) \omega_j \omega_{j'}
     \omega_i + \cdots ,
$$
where $T'_{j\neq 0}$ denotes the MFPT from $t_j$ to $s$ and $T'_0$ is set to
zero.
The first term corresponds to the transition to $t_i$ in a single step,
the second term to a round trip via $t_{j\neq i}$ or staying at $s$ 
followed by the transition to $t_i$, and so on. The infinite sum can easily be 
evaluated which yields
\begin{equation}\label{formal_T}
T_i = {\omega_i}^{-1} \left( 1 + \sum_{j\neq i} \omega_j T'_j\right) \ .
\end{equation}

The configuration space of the Ising spin chain has a tree structure. So
we can make use of the formula in Eq.~(\ref{formal_T}). 
Take a spin state with a boundary domain of $m$ up spins as $s$ 
in Fig.~\ref{fig3}. It is connected to spin states with boundary domains
of $n$ down spins ($n=1,\ldots,m-1$) with the transition probabilities
$p_{m,n}$, which leads to
\begin{equation}\label{T+-}
T^{+-}_m = p_{m,m}^{-1} \left(1 + \sum_{n=1}^{m-1} p_{m,n} T^{-+}_n\right) 
\ .
\end{equation}
Likewise, one also obtains that
\begin{equation}\label{T-+}
T^{-+}_m = q_{m,m}^{-1} \left(1 + \sum_{n=1}^{m-1} q_{m,n} T^{+-}_n\right) 
\ .
\end{equation}
After lengthy but straightforward calculations, the recursion relations
can be solved exactly to yield
\begin{eqnarray}
T^{+-}_m &=& (3-2M) + (3M-2)\,\kappa^{m-2} \\
T^{-+}_m &=& ((3M-2)/M)\, \kappa^{m-1} - 1 
\end{eqnarray}
for $m\geq 2$ and $T^{+-}_1 = M-1$ and $T^{-+}_1 = 1$. Recall
that $\kappa = M/(M-1)$.
The time scales increase {\em exponentially} with $m$.

\subsection{Relaxation time}
Consider an arbitrary spin configuration $|\bfsigma\rangle$ with
$l$ domain walls at sites $\{m_1+1/2,\ldots,m_l+1/2\}$ with $m_i<m_j$ for
$i<j$. The spin state has a boundary domain of size $m_1$ initially.
The spin system loses the memory of the initial state when all spins
flip at least once. Note that $\sigma_G$ is the last 
spin to flip. 
So, the characteristic relaxation time is given by the time 
at which $\sigma_G$ flips for the first time.
It can flip when all spins align ferromagnetically, 
which requires that spins $\sigma_{n\leq m_l}$ align,
which also requires that spins $\sigma_{n\leq m_{l-1}}$ align, and so on.
Therefore the relaxation time is given by 
$T = \sum_{a=1}^l T_{m_a}^{\pm\mp}+T_G^{\pm\mp}$.
For example, the relaxation time for a spin state $|+-++\rangle$
is given by $T = T_2^{+-} + T_3^{-+} + T_4^{+-}$.

Since $T_m^{\pm\mp}$ increases exponentially in $m$, the sum is dominated
by the last term $T_G^{\pm\mp}$ for all spin states. Therefore we conclude
that the characteristic relaxation time averaged over all states scales as
$T \sim T_G^{+-} \sim T_G^{-+}$, which gives
$T \sim \kappa^G$, i.e.\ the important formulas in Eqs.~(\ref{T_G})
and (\ref{T_N}).

\subsection{Autocorrelation}
In this subsection, we derive the scaling laws for 
the autocorrelation function $C_{\bfsigma}(t)$. It measures the strength
of the memory of the initial state $|\bfsigma\rangle$ after time $t$.
The spin system loses the memory as more and more spins fluctuate.
Due to the hierarchical nature of the spin dynamics, the spin fluctuations
grow from one boundary of the chain, namely, from $\sigma_1$.
So, it is useful to define a length scale $\xi(t)$ which is determined 
by the condition that $\sigma_n(t)=\sigma_n(0)$ for $n>\xi$
and $\sigma_{\xi}(t)\neq \sigma_{\xi}(0)$, where $\bfsigma(t)$ denotes
the spin state at time $t$. All spins at sites $n\leq \xi$ have flipped
at least once up to $t$. For this reason, we will call those sites
the {\em perturbed domain}, and $\xi$ the {\em perturbed domain size}.
Roughly speaking, $\xi(t)$ is the maximum size of the boundary domain
up to time $t$.

First, consider the antiferromagnetically ordered state $|A1\rangle$
defined in Eqs.~(\ref{A1state}) and (\ref{A1stateI}). One obtains the same
results for the state $|A2\rangle$. 
It is the linear superposition of $(M-1)^{G/2}$ states, in which 
the random walker is located at 
nodes $|\cdots 0y_30y_1\rangle$, each of which has the degree $K=M-1$. 
Hence, its stationary state probability is given by 
\begin{equation}\label{P_A1}
P_{A1}^\infty \sim (M-1)^{G/2+1}/{\cal N} \sim r_A^{-G}
\end{equation}
with the chain length $G$ and $r_A = M / \sqrt{M-1}$.

The state $|A1\rangle$ has the highest density of domain walls. In
such a state, the perturbed domain grows by removing the domain walls
successively. So, the perturbed domain size reaches $\xi$ after the time
scale $\tau_\xi \sim \sum_{n<\xi} T_n^{\pm\mp} \sim \kappa^\xi$. 
Note that the time scale $t_\xi$ is of the same order of
magnitude as the relaxation time scale of the spin chain of length $\xi$.
It implies that the spins $\sigma_\xi\cdots \sigma_1$ in the perturbed
domain are in the stationary state, while those spins outside the perturbed
domains are frozen at that time scale.
Therefore, $C_A(t_\xi)$ is given by the stationary state
probability for the antiferromagnetic state in the chain of length $\xi$, 
that is $C_{A1}^\infty$ in Eq.~(\ref{P_A1}) with $G$ replaced by $\xi$ to
yield $C_{A1}(t_\xi) \sim r_A^{-\xi}$.
Eliminating $\xi$ in $t_\xi$ and $C_{A1}(t_\xi)$, we obtain the power-law
decay as written in Eq.~(\ref{C_A}). 

\begin{figure}
\includegraphics*[width=\columnwidth]{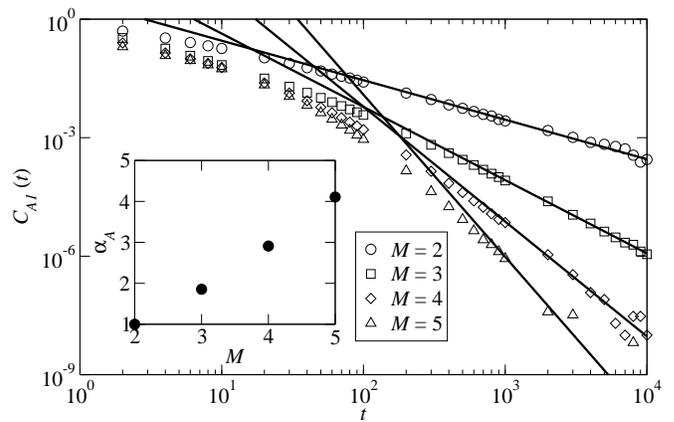}
\caption{Numerical results for $C_{A1}(t)$ are represented by the symbol 
plots for $M=2,3,4,5$ in the Ising spin chain of length $G=100$. 
The solid lines have the slope given by $\alpha_{A1}(M) =\ln
(M/\sqrt{M-1})/\ln \kappa$. The inset shows the plot of $\alpha_{A1}$ vs.
$M$.}\label{fig4}
\end{figure}

We confirmed the analytical results with numerical simulations of the
Ising spin chain. Starting from the initial state $|\cdots +-+-\rangle$,
a stochastic time evolution is generated using the transition rules in
Eqs.~(\ref{pmn}) and (\ref{qmn}) and $C_{A1}(t)$ is measured and averaged
over independent runs. In Fig.~\ref{fig4}, the numerical results are 
presented. They are consistent with the analytic results. 

For the ferromagnetic states $|P\rangle$, one can apply a
similar scaling argument with a little care. 
It is a linear superposition of $(M-1)^G$ states, in which the random walker 
is located on peripheral nodes with degree $M-1+G$. So, its stationary
state probability is given by 
\begin{equation}\label{P_P}
P_P^\infty = (M-1+G) (M-1)^G / {\cal N} \sim G \kappa^{-G} \ .
\end{equation}
The state does not contain any domain walls. So in the beginning 
it evolves quickly creating domain walls into one of states
$\{|\eta\rangle\}$ with $\eta=1,\cdots,G$ with the transition probability 
$p_{G,\eta} \sim 1/G$, where $|\eta\rangle$ denotes a state with the 
domain wall at $\eta+1/2$, i.e., $\sigma_{n>\eta}=+$ and 
$\sigma_{\eta}=-$. After this, the boundary domain growth takes place
for each $|\eta\rangle$ independently. 
After a time scale $t\sim \kappa^\xi$, the spins $\sigma_\xi\cdots\sigma_1$
in the state $|\eta=\xi\rangle$ reaches the stationary state with the 
probability for them to be in the ferromagnetic up state 
is given by $P_P^\infty$ in Eq.~(\ref{P_P}) with $G$ replaced by $\xi$,
i.e., $\xi r_P^{-\xi}$. Therefore the value of the autocorrelation 
function is given by  $C_P(t_\xi) \sim \xi r_P^{-\xi} / G$ 
where $1/G$ is the transition probability from $|P\rangle$ to 
$|\eta=\xi\rangle$. 
Eliminating $\xi$ using $t_\xi \sim \kappa^\xi$, we obtain the result
in Eq.~(\ref{C_P}) in the leading order. 

Analogously, the state $|H\rangle$ evolves into one of the states
$\{|\zeta\rangle\}$ with $\zeta=0,\cdots,G$, 
where $|\zeta\rangle$ denotes a state with the domain wall
at $\zeta+1/2$, i.e., $\sigma_{n>\zeta}=-$ and 
$\sigma_\zeta=+$. In this case, however, the transition probability 
$q_{G,\zeta}\sim (M-1)^\zeta$ increases exponentially with $\zeta$. 
Hence we can ignore the other states except for 
the state with $\zeta=G$, that is $|P\rangle$.
Therefore, the autocorrelation function $C_H(t)$ for $|H\rangle$ is
given by $C_P(t-2)$ multiplied by the transition probability $p_{G,G}$ from 
$|P\rangle$ to $|H\rangle$, which results in Eq.~(\ref{C_H}).

\begin{figure}
\includegraphics*[width=\columnwidth]{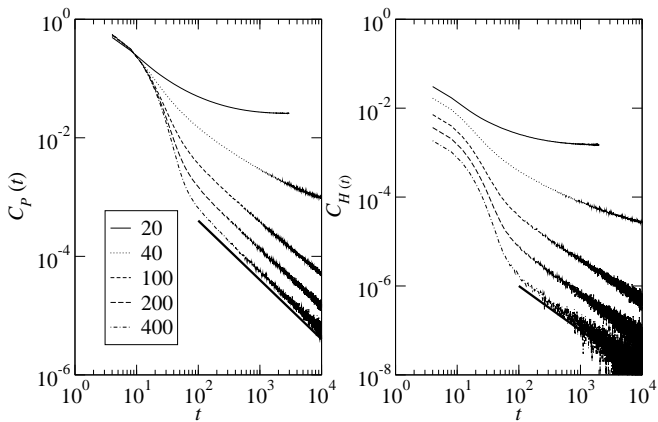}
\caption{Numerical results for $C_P(t)$ and $C_H(t)$ for different 
values of $G$ (as listed in the inset) with $M=5$. The solid lines have
the slope $-1$.}\label{fig5}
\end{figure}
The scaling behavior of $C_P(t)$ and $C_H(t)$ is also confirmed via 
the numerical simulations.
In Fig.~\ref{fig5}, we show a plot of the autocorrelation function
evaluated in the Ising spin chain of length $G\leq 400$ with $M=5$. As $G$
increases, the decay follows the power law in $t$ with the exponent $-1$.
We also checked that the power-law scaling regime overlaps in the plots of
$GC_P(t)$ and $G^2C_H(t)$ vs $t$.

It is easy to generalize the argument for the autocorrelation function to
an arbitrary state $|\bfsigma\rangle$ whose stationary state probability 
scales as $P_{\bfsigma}^\infty \sim r^{-G}$. 
Since the perturbed domain size grows
in time as $\xi \sim \ln t / \ln \kappa$, the value of $C_{\bfsigma}(t)$
at $t\simeq \kappa^\xi$ is given by the stationary state probability
for the spin configuration $\sigma_\xi \cdots \sigma_1$ in the chain of
length $\xi$, i.e., $C_{\bfsigma}(t\simeq \kappa^\xi) \simeq r^{-\xi}$.
Eliminating $\xi$, one obtains that $C_{\bfsigma}(t) \sim t^{-\alpha}$
with a state-dependent exponent $\alpha = \ln r / \ln
\kappa$~\cite{comment}. The stationary state distribution is determined 
by the degree distribution.  Therefore, we conclude that the
non-universality (i.e.\ state dependence) of the decay exponent is a
consequence of the broad distribution of the degree in the underlying
network.

\section{Ultrametric diffusion}\label{sec5}
In the preceding sections it turned out that the origin for the
power-law decay of the autocorrelation functions is the hierarchical
organization of the configuration space; the spins~(or the random
walker) overcome the dynamic barriers successively expanding the
number of accessible configurations. We note that this phenomenon is
very similar to the one observed in the diffusion in an ultrametric
space~\cite{OgielskiS85,PaladinMdeD95}. In this section, we 
compare ultrametric diffusion with the random walk problem
we have studied in this paper.

Consider a dynamical system with $N$ states $a=1,2,\ldots,N$.  The
system in state $a$ may perform transitions to any other state $b$
with a transition probability $w_{ab}$. One can define the distance
between two states as $d_{ab}=1/w_{ab}$ and thus provide the state
space with a metric. If the transition probabilities satisfy the
relation $1/w_{ab} \le \sup ( 1/w_{ac};1/w_{bc})$ for all $a$, $b$,
and $c$, the corresponding metric is called an {\it ultrametric}
and the state space is an ultrametric space.

\begin{figure}
\includegraphics*[width=\columnwidth]{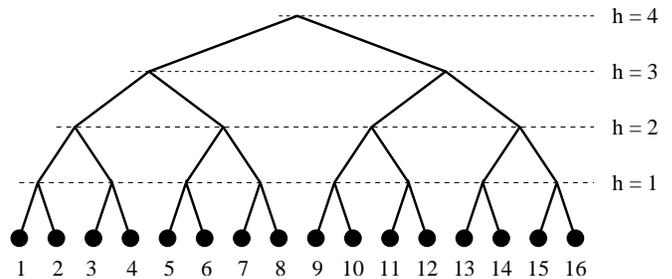}
\caption{Ultrametric space of 16 states.}\label{fig6}
\end{figure}

The simplest example of an ultrametric space is represented by a
rooted tree generated as follows: We start from a single vertex at the
$R$th hierarchy and branch $B$ vertices in the next $(R-1)$th
hierarchy. Each of them branches into $B$ vertices.  It is repeated until
one has $N = B^R$ vertices at the zeroth or bottom level. One then
associates the vertices at the bottom level with the $N$
states. The transition probabilities between two states are assigned
to $w_{ab}=e^{-d\Delta}$, where $\Delta>0$ is a constant and $d$ is
the {\em hierarchical distance} between them, namely the hierarchy
level of their common ancestor at the lowest level. It is easy to see
that the transition probabilities satisfy the ultrametric relation,
and thus an ultrametric space of $N$ states is obtained.  As an
example, we illustrate in Fig.~\ref{fig6} the rooted tree with $R=4$
and $B=2$ for an ultrametric space of $N=16$ states.  In this example,
two states $1$ and $7$ have the common ancestors at the hierarchy
level $h=3$ and $4$, hence $w_{1,7} = e^{-3\Delta}$, while
$w_{1,9}=e^{-4\Delta}$.

The autocorrelation function can be calculated exactly, see e.g., 
Ref.~\cite{OgielskiS85}. The exact result is also understood
with a simple scaling argument. 
Suppose that the system is in a state $a$ initially. 
Since the transition probability to a state at the hierarchical distance
$\xi$ is given by $w=e^{-\Delta\xi}$ and there are ${\cal O}(B^\xi)$
such states, it takes $t_\xi \sim (B^\xi e^{-\Delta \xi})^{-1}$ time
steps for the system to reach one of the states within the hierarchical
distance $\xi$. Hence, the autocorrelation function at time $t\sim t_\xi$
is given by $C(t_\xi) \sim B^{-\xi}$.
Eliminating $\xi$, one obtains that the autocorrelation function decays 
algebraically as $C(t) \sim t^{-\alpha}$ with 
$\alpha = \ln B / (\Delta - \ln B)$. 
The power-law decay is valid for $\Delta > \ln B$, while the dynamics is
unstable for $\Delta<\ln B$. At the marginal case, a stretched exponential 
decay $P(t) \sim e^{-(\ln B) t^{1/\gamma}}$ may occur when
the transition probability decreases as $w(d) \sim d^{-\gamma} 
e^{-(\ln B)d}$ with the hierarchical distance~\cite{OgielskiS85,PaladinMdeD95}.

Comparing the phenomenology, it is clear that the diffusion in 
the hierarchical network is essentially the same as the ultrametric 
diffusion. In both processes, the relaxation takes place by overcoming
dynamic barriers successively and increasing associated length scale.
The length scale corresponds to the perturbed domain size $\xi(t) \sim \ln t
/ \ln \kappa$ in the former, 
and to the hierarchical distance $\xi(t) \sim \ln t / \ln (\Delta - \ln B)$
in the latter. 
The length scale grows logarithmically in time, which is a consequence of
the exponential increase of the dynamical barrier height. 

Note, however, that the diffusion in the hierarchical network is
not the ultrametric diffusion in a strict sense since 
the ultrametric relations are not valid.
The configuration space of the Ising spin system has a tree structure 
with {\em all vertices} corresponding to physical states.
The ultrametricity would hold only if vertices at the bottom hierarchy 
would represent physical states, see Figs.~\ref{fig2} and \ref{fig6}.
Such a difference does not modify the ultrametric nature of the relaxation 
from the state $|A1\rangle$ and $|A2\rangle$, which are located at the end 
branch in the configuration space. 
On the other hand, the relaxation from $|H\rangle$ and $|P\rangle$,
which are in the center of the configuration space tree, are influenced
by the non-ultrametricity. It is reflected in the $G^{-1}$ and $G^{-2}$
factors in the autocorrelation functions $C_{P}(t)$ and $C_{H}(t)$,
respectively.
Pseudo-ultrametric diffusion is also observed for
the random walks on a tree structure~\cite{Schreckenberg} and 
on the one-dimensional lattice with hierarchically distributed
dynamic barriers~\cite{HubermanK85}.

\section{Summary}\label{sec6}
In summary, we have studied the random walk problem on the
hierarchical network.  The random walk problem on the network of
$N=M^G$ nodes is mapped to a specially constrained dynamics of a
$M$-state Potts spin chain of length $G$. Using the symmetry property,
it is further mapped to a specially constrained dynamics of an Ising
spin chain.  From the analysis of the MFPT, it is shown that the
characteristic relaxation time scales as $T \sim \kappa^G \sim N^z$
with $\kappa = M/(M-1)$ and $z=\ln \kappa/\ln M$. It is also shown
that the autocorrelation function decays algebraically in time as
$C_{\bfsigma}(t)
\sim t^{-\alpha_{\bfsigma}}$ for $t\ll T$ with a non-universal 
(i.e.\ state-dependent) exponent $\alpha_{\bfsigma}$. 
The power-law scaling behavior is closely related to the ultrametric diffusion.
The exponent is given by $\alpha_{\alpha} = \ln r_{\bfsigma} / \ln \kappa$
for a state $\bfsigma$ whose stationary state probability is
$P_{\bfsigma}^\infty \sim r_{\bfsigma}^{-G}$. The stationary state
probability is determined from the degree of the corresponding nodes in the
network. The broad distribution of the degree gives rise to
the non-universality (state-dependency) of the decay exponent.

The power-law decay of the autocorrelation functions appears in
marked contrast to the stretched-exponential decay in random
networks~\cite{Bray&Rodgers88} and in the small-world
networks~\cite{AlmaasKS03}. In order to investigate the origin of the
emergence of the power-law scaling, we have also studied the random
walks on the hierarchical networks with $M=2$.  At $M=2$, the
hierarchical network is not scale-free any more.  Nevertheless, we can
use the same mapping to the Ising spin system with the configuration
space of the same tree structure.  So we can obtain the scaling
behaviors of the relaxation time $T$ using Eqs.~(\ref{T+-}) and
(\ref{T-+}), and of the autocorrelation function using the same
scaling arguments: The relaxation time scales as $T\sim G^{\ln 2} \sim
N^1$. And the autocorrelation functions decay algebraically in time
with the {\em universal} (state-independent) exponent, i.e., $\alpha_H
= \alpha_P = \alpha_A = 1$.  For $M=2$, the corresponding spin
dynamics has the spin up-down symmetry.  So, $C_H(t)$ and $C_P(t)$
decay in the same way as $C_H(t) = C_P(t) \sim t^{-1} / G$ with the
same dependency on $G$. Finally, the scaling behavior of the
relaxation time and the autocorrelation functions was confirmed
numerically.

Comparing the results for $M=2$ and $M>2$, we conclude that the
power-law scaling behavior of the relaxation time and the
autocorrelation functions has its origin in the tree structure of
the spin configuration space as shown in Fig.~\ref{fig2}. 
We also conclude that the non-universality of
the decay exponent for $M>2$ results from the scale-free degree
distribution.

The hierarchical network itself does not have a tree structure.
But, after the mapping, the random walk problem on the network
reduces to that on the tree structure. In general, it is not
known {\em a priori} whether such a mapping exists for an arbitrary network.
It would be interesting to study the random walk problem on general 
networks in order to scrutinize the robustness of the power-law scaling
behavior and the effect of the scale-free degree distribution
on the relaxation dynamics. Such work is actually in progress.

We note that the very slow relaxation dynamics of the Ising chain
representation of the random walk problem on the hierarchical network
is due to the severe constraints of the dynamics imposed by the
restrictions for possible transitions. Constrained dynamics in
otherwise (for instance thermodynamically) very simple models lead
quite frequently to a slow or {\it glassy} dynamics
\cite{constrained-review}, for which reason kinetically constrained 
models are often used as models for the dynamics in glasses and spin
glasses. It is interesting to note that such a model also occurs in
the context of diffusion in complex networks as we have demonstrated
in this work.

\acknowledgments
This work was supported by the Deutsche Forschungsgemeinschaft~(DFG),
by the KOSEF Grant No. R14-2002-059-01000-0 in the ABRL program and by
the European Community's Human Potential Programme under contract
HPRN-CT-2002-00307, DYGLAGEMEM.


\begin{references}
\bibitem{comment} It is assumed that the constant $r$ does not depend on the
length scale. So, the power law scaling in time will be observed for the
states where spins are distributed rather uniformly. 
A complex behavior would emerge otherwise.
\end{references}
\end{document}